\documentstyle{article}
\newcommand{\mbs}[1]{\mbox{\scriptsize{#1}}}

\renewcommand{\(}{\left (}
\renewcommand{\)}{\right )}
\newcommand{\beq}{\begin{equation}}
\newcommand{\eeq}{\end{equation}}
\newcommand{\beqa}{\begin{eqnarray}}
\newcommand{\eeqa}{\end{eqnarray}}
\input{psfig}
\begin{document}
\begin{titlepage}
\begin{flushright}
LU TP 97-2\\
March 6, 1997 \\
\end{flushright}

\vspace{0.8in}
\LARGE
\begin{center}
{\bf A Potts Neuron Approach to\\ Communication Routing}\\
\vspace{.3in}
\large
Jari H\"akkinen\footnote{jari@thep.lu.se},
Martin Lagerholm\footnote{martin@thep.lu.se},
Carsten Peterson\footnote{carsten@thep.lu.se} and
Bo S\"{o}derberg\footnote{bs@thep.lu.se}\\
\vspace{0.05in}
Complex Systems Group, Department of Theoretical Physics\\
University of Lund, S\"{o}lvegatan 14A, SE-223 62 Lund, Sweden\\
\vspace{0.15in}

Submitted to {\it Neural Computation}

\end{center}
\vspace{0.2in}
\normalsize

Abstract:

A feedback neural network approach to communication routing problems
is developed with emphasis on {\it Multiple Shortest Path} problems,
with several requests for transmissions between distinct start- and
endnodes.
The basic ingredients are a set of Potts neurons for each request,
with interactions designed to minimize path lengths and to prevent
overloading of network arcs. The topological nature of the problem is
conveniently handled using a propagator matrix approach. Although the
constraints are global, the algorithmic steps are based entirely on
local information, facilitating distributed implementations.
In the polynomially solvable single-request case the approach reduces
to a fuzzy version of the {\it Bellman-Ford} algorithm.
The approach is evaluated for synthetic problems of varying sizes and
load levels, by comparing with exact solutions from a branch-and-bound
method.
With very few exceptions, the Potts approach gives legal solutions of
very high quality. The computational demand scales merely as the
product of the numbers of requests, nodes, and arcs.

\end{titlepage}

\newpage

\newpage

\section*{Introduction}

Communication routing resource allocation problems are becoming
increasingly relevant given the upsurge in demand of internet and
other telecommunication services.  One such problem amounts to
assigning arcs (links) in a connected network to requests from start-
to endnodes, given capacity constraints on the links, such that a
total additive cost (path-length) is minimized. For a review of
notation and existing routing techniques, see e.g. ref. \cite{bert}. A
relatively simple routing problem, with only one request at a time, is
the Shortest Path Problem ({\bf SPP}), which can be solved exactly in
polynomial time using e.g. the Bellman Ford ({\bf BF)} algorithm
\cite{bert}. The Multiple Shortest Path Problem ({\bf MSPP}), where
links are to be allocated simultaneously to several requests, is more
difficult. There exists, to our knowledge, no method that yields exact
solutions to this problem in polynomial time.

In this paper we address the MSPP using feedback Potts neural
networks, which have proven to be powerful in other resource
allocation problems, with \cite{lager} or without \cite{gis2} a
non-trivial topology.  For each request we assign a Potts network,
with units encoding which links to be used by that request.
Appropriate energy terms are constructed in terms of the Potts neurons
to minimize total path lengths and to ensure that capacity constraints
are not violated. Mean field ({\bf MF}) equations are iterated using
annealing to minimize the total energy. In contrast to earlier usage
of Potts encoding and MF annealing \cite{pet2,gis2,lager} where
global objective functions are minimized, here each node minimizes
its own local energy.
%
%

For the case of a {\em single} request, the Potts MF approach reduces
in the zero temperature limit to the BF algorithm; hence our approach
contains this standard algorithm as a special case.

For each request the Potts MF network \cite{pet2} defines a
``fuzzy'' {\em spanning tree}\footnote{A subgraph connecting all
nodes without loops.}, rooted at the endnode.  In order to project out the
(fuzzy) path from the startnode in this subgraph, and to keep track
of the paths in general, we utilize a propagator matrix formalism
following ref. \cite{lager}.  The computation of the propagator
requires matrix inversion; fortunately this can be done using an
iterative procedure with low computational cost.

As in a previously considered airline crew scheduling problem
\cite{lager}, proper preprocessing is employed to identify independent
subproblems, in order to reduce the problem complexity.

Despite the existence of global constraints, the implementation
of the approach is truly local -- when updating the MF equations
for a particular node, only information residing at neighbouring
nodes is needed.

The approach is gauged by an exact branch-and-bound ({\bf BB})
algorithm on a set of synthetic but challenging test problems, showing
an excellent performance of the Potts MF approach, with a CPU
consumption per request scaling merely like $NN_L$, where $N$ is the
number of nodes and $N_L$ the number of links in the network. The
method is also very robust with respect to parameters. Due to the
excessive demand for CPU resources by the reference BB method, our
comparisons are limited to fairly low problem sizes.  However, there
are no indications that the Potts method be less efficient for larger
problems.

\section*{The Multiple Shortest Path Problem}

\begin{figure}
\begin{center}
  \mbox{\psfig{figure=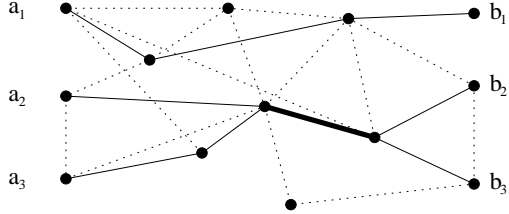,width=6.7cm}}
\end{center}
\caption{Example of a solution to a 3-request problem.  Dotted lines
  represent unused links, and full lines links used by the requests.}
\label{msppfig}
\end{figure}

A Multiple Shortest Path Problem (MSPP) is defined by specifying the
following:
\begin{itemize}
\item A connected network of $N$ nodes and $N_L$ links (arcs).
\item For each arc $ij$, a cost (arc-length) $d_{ij}$ and a capacity
	$C_{ij}$.
\item A set of $N_R$ transmission requests $r$, each defined by a
	startnode $a_r$ and an endnode $b_r$.
\end{itemize}
The task is then to assign to each request a connected loop-free path
of arcs from the startnode to the endnode. This is to be done such
that the total length of those paths is minimized, without the load on
any arc exceeding its capacity, with the load defined as the number of
requests sharing it. A 3-request problem example is given in
fig.~\ref{msppfig}.

All problems of this kind are not solvable; a reliable algorithm should
be able to recognize and signal a failure, to enable proper measures to
be taken.

\section*{The Bellman-Ford Algorithm in the Mean Field Language}

Prior to dealing with the MSPP, we revisit the simpler SPP and
demonstrate how the BF algorithm can be recast in a Potts MF
language. This formulation will then be the starting point for
designing a Potts MF approach to MSPP.

In the SPP there is only a single request, from $a$ to $b$, and the
capacity constraints are irrelevant. The task simply is to find the
shortest path from $a$ to $b$. In the BF algorithm \cite{bert} this is
achieved by minimizing the path-lengths $D_i$ from every node $i$ to
$b$, by iterating
\beq
\label{BF}
	D_i \to \min_j (d_{ij}+D_j) , \;\;\; i \ne b.
\eeq
and keeping track of the involved links $ij$. $D_b$ is fixed to zero
by definition.  The resulting solution defines a spanning tree rooted
at $b$.  In particular, $D_a$ is determined, and the minimal path from
$a$ to $b$ is easily extracted from the spanning tree.

If no link exists from node $i$ to $j$, $d_{ij}$ could formally
be defined to be infinite; in practice it is more convenient to
restrict $j$ in eq.~(\ref{BF}) to the actual {\em neighbours} of $i$,
reachable via an arc from $i$.

Eq.~(\ref{BF}) can be rewritten as
\beq
\label{BF_1}
	D_i = \sum_j v_{ij} E_{ij} \equiv \sum_j v_{ij} (d_{ij}+D_j)
\eeq
in terms of a {\em Potts spin} ${\bf v}_i$ for every node $i \neq b$,
with components $v_{ij}$ taking the value 1 for the $j$ with the
smallest {\em local energy} $E_{ij}$, and $0$ for the others
(winner-takes-all).  Note the distinct philosophy here: each node $i$
minimizes its own local energy $D_i = \min_j E_{ij}$, rather than all
nodes striving to minimize some global objective function.

A {\em mean field} (MF) version of eq.~(\ref{BF_1}) is obtained by
using for ${\bf v}_i$ its thermal average in the MF approximation,
defined by
\beq
\label{potts}
	v_{ij} = \frac{e^{-E_{ij}/T}}{\sum_ke^{-E_{ik}/T}}
\eeq
where $j$ and $k$ are neighbours of $i$, and $T$ is an artificial
temperature.  Note that each {\em Potts MF neuron} ${\bf v}_i$ obeys
the normalization condition
\beq
\sum_j v_{ij} = 1
	\label{potts_cond}
\eeq
allowing for a probabilistic interpretation of the components.

At a non-zero temperature, iteration of eqs.~(\ref{BF_1},\ref{potts})
can be viewed as a fuzzy implementation of the BF algorithm, while in
the $T \to 0$ limit the neurons are forced {\em on-shell}, i.e.
$v_{ij} \to 1$ (for the minimizing~$j$) or 0 (for the rest), and
proper BF is recovered.

Given this obvious neural recast of the BF algorithm in terms of Potts
neurons, it is somewhat surprising that non-exact neural approaches
based on Ising spins have been advocated in the literature~\cite{thomo}.

In addition, the MF Potts algorithm considered here exhibits a close
electrostatic analogy in terms of Kirchhoff's laws on a graph
\cite{jari1}.

\section*{The Potts Mean Field Approach to MSPP}

The Potts MF formulation of the Bellman-Ford algorithm for SPP
(eqs. (\ref{BF_1},\ref{potts})) is a suitable starting point for
approaching MSPP.

We will stick with the philosophy inherited from BF of focusing on
independent {\em local energies}, in contrast to what has become
standard when using feedback neural networks for resource allocation
problems. This represents a novel strategy.

Thus, we introduce a separate Potts system, $\{v_{ij}^r\}$, for each
request $r$, with basic local energies $E_{ij}$ as before representing
distances to the endnode $b_r$.  In addition, we will need energy
terms $E^{\mbs{load}}_{ij}$ for the load-constraint, to be discussed
later; this introduces an interaction between the Potts systems.

This formulation introduces the possibility of undesired loop
formation, since forming a loop might induce less energy penalty than
violating a load constraint.  As will be discussed below such loops
can be suppressed by a suitable penalty term, $E^{\mbs{loop}}_{ij}$,
and by adding a possibility for each proper node to connect, via an
artificial {\em escape link}, to an artificial {\em escape node}, for
each request connecting directly to the endnode.
%
This enables a ``give-up'' state for an unresolvable situation,
signaled by some path containing the escape node. The cost for
``giving up'' must be larger than that for any legal path.  Therefore
the length of each escape link is set to the sum of the lengths of the
proper links, while the corresponding capacity is chosen large enough
to be able to host all the requests.

In order to terminate the path for a request $r$, its endnode must be
a {\em sink} for the corresponding Potts system. Consequently, there
will be no Potts neuron ${\bf v}^r_{b_r}$ associated with it.

In order to construct appropriate penalty terms, a propagator matrix
will be used. This technique has proved to be a powerful tool in
neural optimization for problems with a non-trivial topology
\cite{lager}.  In particular, it will be crucial for extracting
properties of the fuzzy paths defined by the MF approach at finite
$T$.

\subsection*{Path Extraction and the Propagator}

\begin{figure}
\mbox{
\psfig{figure=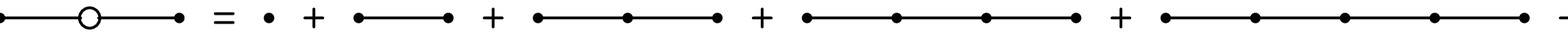,width=14.8cm,height=.6cm}
}
\vspace{0.2in}
\caption{Expansion of the propagator $P^r_{ij}$ in terms of $v^r_{kl}$
	(---) and intermediate nodes ($\bullet$).}
\label{fig_prop}
\end{figure}

The normalization condition (eq. (\ref{potts_cond})) ensures that
for each request there is precisely one continuation for each node,
although for $T \neq 0$ it is fuzzily distributed over the available
neighbours.
On shell, the path from start- to endnode is trivial to extract -- one
follows the $v^r_{ij} = 1$ path starting from the startnode.  However,
for $T \neq 0$ a more refined path extraction mechanism is needed.
This is provided by a {\em propagator matrix} \cite{lager} ${\bf P}^r$
for each request $r$, defined by:
\beq
\label{prop}
	P^r_{ij} = \left[ \( {\bf 1} - {\bf v}^r \)^{-1} \right]_{ij}
	= \delta_{ij} + v^r_{ij} + \sum_k v^r_{ik} v^r_{kj}
	+ \sum_{kl} v^r_{ik} v^r_{kl} v^r_{lj}
	+ \ldots
\eeq
For a graphical representation, see fig.~\ref{fig_prop}. On shell, it
is easy to see that $P^r_{ij}$ can be interpreted as the {\em number
of paths} from $i$ to $j$ defined by the Potts neurons associated with
$r$; similarly, the elements of the matrix square $(P^r)^2_{ij}$ are related to the
number of arcs used in those paths\footnote
{More precisely, the number of arcs is given by $P^2-P$.}.

Off shell, these interpretations are still valid in a probabilistic
sense.  A probabilistic measure of how much node $i$ participates in
the path $a_r \to b_r$ is given by $P^r_{a_ri} P^r_{ib_r} \equiv
P^r_{a_ri}$, the simplification being due to the identity
$P^r_{ib_r}=1$, following from eq.~(\ref{potts_cond}). This provides
us with a natural path extractor; in particular, we have on shell (in
the absence of loops)
\beq
	P^r_{a_ri} =
	\left\{
	\begin{array}{ll}
	1, & \mbox{if node $i$ appears in the path $a_r \to b_r$} \\
	0, & \mbox{otherwise}
	\end{array}
	\right.
\label{P_Ai}
\eeq
%

\subsection*{Load and Loop Control}

Armed with the propagator formalism, we proceed to set up the penalty
terms, to be added to the energies $E_{ij}$ corresponding to
eq. (\ref{BF_1}).

In order to efficiently avoid complications from self-interactions in
the local energies, independently of $T$, we follow ref. \cite{ohl},
and define the penalty terms based on analyzing the result of setting
one component $v^r_{ij}$ of the neuron ${\bf v}^r_i$ at a time to one,
with the other components set to zero, as compared to a reference
state with all components set to zero.

The total load $L_{ij}$ on a link $ij$ is the sum of contributions
from the different requests,
\beq
	L_{ij} = \sum_r L^r_{ij}.
\label{L}
\eeq
The contribution from request $r$ can be expressed as
\beq
	L^r_{ij} = \frac{ P^r_{a_ri} }{ P^r_{ii} } v^r_{ij}
\label{L_r}
\eeq
where $P^r_{ii}$ corrects for possible improper normalization due
to soft loop-contributions.

The load constraints, $L_{ij} \leq C_{ij}$, define a set of inequality
constraints. In the realm of feedback neural networks, such
constraints have been successfully handled by means of step-functions
\cite{ohl}.
For a given request $r$, the available capacity of the link $ij$ is
given by
\beq
	X \equiv  C_{ij} - L_{ij} + L^r_{ij} \; ,
\eeq
in terms of which an overloading penalty can be defined as
\beq
E^{\mbs{load}}_{ij} =
	\left(1-X\right) \Theta\left(1-X\right)
	+ \left(X\right) \Theta\left(-X\right) \\
\label{E_load}
\eeq
where $\Theta()$ is the Heaviside step-function.  Eq.~(\ref{E_load})
expresses the additional overloading of the link, if it were to be
used by $r$.

The amount of loops introduced by connecting $i \to j$ can be
expressed as
\beq
	Y \equiv  \frac{ P^r_{ji} }{ P^r_{ii} }
\eeq
A suitable loop suppression term is then given by
\beq
	E^{\mbs{loop}}_{ij} = \frac{Y}{1-Y}.
\label{E_loop}
\eeq

The generalization of the local energy in eq.~(\ref{BF_1}) to the
multiple request case now reads, for a particular request $r$,
\beq
\label{BF_2}
	E_{ij} = d_{ij} + D^r_j
	+ \alpha E^{\mbs{load}}_{ij}
	+ \gamma E^{\mbs{loop}}_{ij}
\eeq
with the added terms based on eqs. (\ref{E_load},\ref{E_loop}). The
resulting algorithm allows for a wide range of choices of the
coefficients $\alpha$ and $\gamma$ without severely changing the
performance.

\subsection*{Updating Equations}

All neurons are repeatedly updated, with a slow annealing in $T$.  For
each request $r$ and each node $i$, the corresponding neuron ${\bf
v}^r_i$ is updated according to
\beq
\label{multi_potts}
	v^r_{ij} = \frac{e^{-E_{ij}/T}}{\sum_ke^{-E_{ik}/T}}.
\eeq
with $E_{ij}$ given by eq.~(\ref{BF_2}).  The corresponding length
$D^r_i$ from node $i$ to the endnode is then updated, in the BF
spirit, as
\beq
	D^r_i \to \sum_j v^r_{ij} E_{ij}.
\label{D_soft}
\eeq
The corresponding update of the propagator could in principle be done
using an exact incremental matrix inversion scheme like
Sherman-Morrison \cite{numrec}.  We prefer, though, to let local
changes propagate through the network, in analogy to the update of
$D^r_i$. Thus, only the $i$'th row of ${\bf P}^r$ is updated:
\beq
	P^r_{im} \to \delta_{im} + \sum_j {v^r_{ij} P^r_{jm}}
	\; , \;\; \mbox{for all } m
\label{P_soft}
\eeq
This gives a convergence towards the exact inverse, which turns out to
be good enough.  The advantage of this method is twofold: it is
faster, and all information needed is local to the relevant node $i$
and its neighbours $j$ (assuming each node to keep track of its own
row of ${\bf P}^r$).

Details of the algorithmic steps and the initialization can
be found in Appendix A.

\section*{Test Problems and Explorations}

In order to test the Potts MF method we have generated a set of
synthetic problems.
In doing so we face limitations given by the ability of the exact
reference method (BB) to provide solutions, due to limited computing
power. For this reason our comparisons are restricted to relatively
small systems.

With these limitations in mind, we have tried to choose the test
problems as difficult as possible.  The most important parameters
governing the difficulty of a problem, apart from network size and
connectivity, are the number of requests and the average link
capacity.  In cases where all the links are able to host all the
requests, $C_{ij} \geq N_R$, the problem is separable, and can be
solved by running an independent BF algorithm for each
request. Therefore we choose to work with very tight link bounds.

For each problem, we generate a random connected network, where every
node has at least one path to all other nodes.  To that end, all nodes
are first connected in a random spanning tree. Additional links
(creating loops) are then randomly placed.
Every link is given a random capacity and a random length.
\begin{figure}[htb]
\begin{center}
  \mbox{\psfig{figure=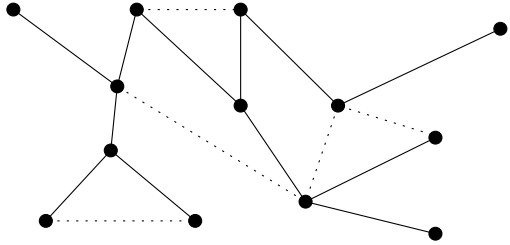,width=6.7cm}}
\end{center}
\caption{An example network with 13 nodes and 17 links. The initial
spanning tree are the solid lines whereas the dotted lines represent the links
creating loops.}
\label{f:typnet}
\end{figure}
Finally a specified number of random requests are generated, in terms
of start- and endnodes.  An example of a generated test problem is
shown in fig~\ref{f:typnet}.

This procedure does not automatically yield a solvable problem, where all
requests can be fulfilled simultaneously without violating any
constraint.  In principle, solvability could be built into the
problem-generator.
Here we adopt another strategy, by creating random problems and
attempting to solve them exactly; those with the solution
not found by BB within a certain amount of CPU time are disregarded.
In principle, this method could introduce a bias towards simple
problems.  However, for the problem sizes considered only a tiny
fraction of the problem candidates are ``timed-out'', except for the
largest problems (last row in Table~\ref{f:typnet}), where about 1/3
are disregarded.

Prior to attacking a problem, a decomposition into independent
sub-problems is attempted, to reduce complexity. The required computer
time for a decomposed problem is dominated by the largest
sub-problem. All sub-problems are solved in the explorations described
below.  The CPU demand of the MF approach scales like $N_R N N_L$.  

The performance of the MF Potts approach is probed my measuring the
excess path length as compared to the BB result,
\beq
	\Delta = \frac
	{ D_{\mbs{MF}} - D_{\mbs{BB}} }
	{ D_{\mbs{BB}} }.
\eeq
where $D_{\mbs{MF}}$ and $D_{\mbs{BB}}$ are the total path lengths
resulting from the two algorithms.
In Table~\ref{res} we show characteristics of the generated test
problems, together with performance measures for the Potts MF
algorithm.  As a measure of the complexity of a problem we use the
entropy, $S$, defined as the logarithm of the total number of possible
configurations, disregarding load constraints.
%
\begin{table}[h]
\begin{center}
\begin{tabular}{|c|c|c|c||c|c|c|}
\hline
Nodes & Links & Requests & $<S>$ & \% legal & $<\Delta>$ & $<$CPU-time$>$ \\
\hline
 5 & 10  & 5  &   14 & 100.0 & 0.003 & 0.1\\
 5 & 10  & 10 &   28 &  99.9 & 0.002 & 0.2\\
10 & 15  & 10 &   28 & 100.0  & 0.004 & 0.4\\
10 & 20  & 10 &   48 & 100.0  & 0.003 & 0.5\\
15 & 20  & 10 &   27 &  99.8 & 0.03  & 0.5\\
15 & 20  & 15 &   40 &  99.9 & 0.06  & 0.7\\
\hline
\end{tabular}
\end{center}
\caption{Results averaged over 1000 problems of each size.
 The entries ``$\%$ legal'' and $<$CPU time$>$ refer to the MF Potts
 approach. The time is given in seconds using DEC Alpha 2000. Typical
 times for the BB method are around 600 seconds.}
\label{res}
\end{table}

Table~\ref{res} indicates an excellent performance of the MF Potts
approach, with respect to giving rise to legal solutions with good
quality, with a very modest computational demand.

\section*{Summary and Outlook}

We have developed a Potts MF neural network algorithm for finding
approximate solutions to the Multiple Shortest Path Problem. The
starting point is a Potts MF recast of the exact Bellman-Ford
algorithm for the simpler single Shortest Path Problem.  This approach
is then extended to the Multiple Shortest Path Problem by utilizing
several Potts networks, one for each request. Complications of
topological nature are successfully handled by means of a convenient
propagator approach, which is crucial for the following issues:
\begin{itemize}
\item The MF approach yields at $T\neq 0$ fuzzy spanning trees, from
     which the propagator is used to extract fuzzy paths.
\item The load-constraints are handled by energy terms involving the
     propagator.
\item Loops are suppressed by energy terms, also based on the propagator.
\end{itemize}
In addition, an auxiliary link to an escape node is introduced for
each proper node, opening up escape paths for unresolvable situations.

The method is local in that only information available from
neighbouring nodes is required for the updates. This attractive
feature, inherited from the BF algorithm, facilitates a distributed
implementation.

The computational demand of the method is modest. The CPU time scales
as $N_RNN_L$. With fixed connectivity this corresponds to $N_RN^2$,
whereas for the worst case of full connectivity it yields $N_RN^3$.

The performance of the algorithm is tested on a set of synthetic
challenging problems, by comparing to exact results from a
branch-and-bound method, for various problem sizes.  The comparison
shows that the Potts MF approach with very few exceptions yields very
good approximate solutions.

The method is presently being generalized to other routing problems.

\appendix
\renewcommand{\thesection}{Appendix \Alph{section}.\setcounter{equation}{0}}
\renewcommand{\theequation}{\Alph{section}\arabic{equation}}
\newcommand{\app}{\newpage\section}


\app{The Potts MF Algorithm}

\subsection*{Initialization}

The initial temperature $T_0$ is first set to $T_0=50$. If the
saturation $\Sigma$,
\beq
 \Sigma \equiv \frac{1}{N_R(N-1)}\sum_{i \neq b_r} {\bf v}_i^2,
\eeq
has changed more than 10\% after all neurons have been 
updated once, then the system is reinitialized with $T_0 \to 2T_0$.

For all nodes (except the end- and escape nodes), the
corresponding  Potts neurons are initialized in accordance 
with the high temperature limit,
i.e.
\beq
	v^r_{i,j} = 1/n_i
\label{v_init}
\eeq
for all $n_i$ neighbours $j$ (including the escape node) of $i$ .
$P^r_{ij}$ and $D^r_i$ are initialized consistently with
eq.~(\ref{v_init}).
%

\subsection*{Iteration}

\fbox{
\begin{minipage}{15cm}
Until $T \leq T_f$ or $\Sigma \geq \Sigma_f$ do:
\begin{itemize}
  \vspace{0.1in}
  \item For every request $r$ do:
     \begin{enumerate}
  	\item For every node $i$ except $b_r$ and the escape node:
  	\begin{enumerate}
		\item Update ${\bf v}^r_i$ (eqs.~(\ref{potts},\ref{BF_2})).
    		\item Update $D^r_i$  (eq.~(\ref{D_soft})).
    		\item Update ${\bf P}^r_i$ (eq.~(\ref{P_soft})).
  	\end{enumerate}
  	\item Update $L_{ij}$.
      \end{enumerate}
  \item Decrease the temperature: $T=kT$.
 \end{itemize}
\end{minipage}
}

Parameters used are: $k=0.90$, $T_f$=0.0001, $\Sigma_f$=0.99999.  For
the energy coefficients in eq. (\ref{BF_2}), we have consistently used
$\alpha=1$ and $\gamma=5$.

\end{document}